\documentclass[12pt,preprint]{aastex}
\usepackage{natbib}

\begin{document}

\title{A Targeted Search for Point Sources of EeV Neutrons}

\author{The Pierre Auger Collaboration\altaffilmark{\dag}} 
\altaffiltext{\dag}{Pierre Auger Collaboration, Av. San Mart\'{\i}n Norte 306, 5613 Malarg\"ue, Mendoza, Argentina; \url{www.auger.org}}

\author{
\begin{small}
A.~Aab$^{42}$, 
P.~Abreu$^{65}$, 
M.~Aglietta$^{54}$, 
M.~Ahlers$^{95}$, 
E.J.~Ahn$^{83}$, 
I.~Al Samarai$^{29}$, 
I.F.M.~Albuquerque$^{17}$, 
I.~Allekotte$^{1}$, 
J.~Allen$^{87}$, 
P.~Allison$^{89}$
% ...
\\ \vspace{0.5cm}
$^{1}$ Centro At\'{o}mico Bariloche and Instituto Balseiro (CNEA-UNCuyo-CONICET), San 
Carlos de Bariloche, 
Argentina \\
% ...
$^{17}$ Universidade de S\~{a}o Paulo, Instituto de F\'{\i}sica, S\~{a}o Paulo, SP, 
Brazil \\
% ...
$^{29}$ Institut de Physique Nucl\'{e}aire d'Orsay (IPNO), Universit\'{e} Paris 11, CNRS-IN2P3, 
Orsay, 
France \\
% ...
$^{42}$ Universit\"{a}t Siegen, Siegen, 
Germany \\
% ...
$^{54}$ Osservatorio Astrofisico di Torino  (INAF), Universit\`{a} di Torino and Sezione INFN, 
Torino, 
Italy \\
% ...
$^{65}$ Laborat\'{o}rio de Instrumenta\c{c}\~{a}o e F\'{\i}sica Experimental de Part\'{\i}culas - LIP and  
Instituto Superior T\'{e}cnico - IST, Universidade de Lisboa - UL, 
Portugal \\
% ...
$^{83}$ Fermilab, Batavia, IL, 
USA \\
% ...
$^{87}$ New York University, New York, NY, 
USA \\
% ...
$^{89}$ Ohio State University, Columbus, OH, 
USA \\
% ...
$^{95}$ University of Wisconsin, Madison, WI, 
USA \\
% ...
(\ddag) Deceased \\
(a) Now at Konan University \\
(b) Also at the Universidad Autonoma de Chiapas on leave of absence from Cinvestav \\
(c) Now at NYU Abu Dhabi \\
% last updated:	01.04.2014 
\end{small}
}

\begin{abstract}

\noindent A flux of neutrons from an astrophysical source in the
Galaxy can be detected in the Pierre Auger Observatory as
an excess of cosmic-ray air showers arriving from the direction of the source.
To avoid the statistical penalty for making many trials, classes of
objects are tested in combinations as nine ``target sets'', in addition
to the search for a neutron flux from the Galactic Center or from the
Galactic Plane.  Within a target set, each candidate source is
weighted in proportion to its electromagnetic flux, its exposure to
the Auger Observatory, and its flux attenuation factor due to neutron
decay.  These searches do not find evidence for a neutron flux from
any class of candidate sources. Tabulated results give the combined
p-value for each class, with and without the weights, and also the
flux upper limit for the most significant candidate source within each
class. These limits on fluxes of neutrons significantly constrain
models of EeV proton emission from non-transient discrete sources in
the Galaxy.

\end{abstract}


%%%%%%%%%

\section{Introduction}
\label{introduction}

\noindent The Pierre Auger Observatory measures cosmic-ray
air showers near 1~EeV and higher energies (1~EeV = $10^{18}$~eV).
An air shower produced by a neutron is indistinguishable from an
air shower produced by a proton.  Unlike protons and other
nuclei, neutrons are not deflected by magnetic fields in the Galaxy,
so their arrival directions point back to their sources.  A flux of
neutrons from a single direction can be detected as an excess of air
showers arriving from that direction within the angular resolution of
the Observatory.  The mean decay path length for a neutron of energy
$E$ (measured in EeV) is $9.2E$ kpc.  Above 1~EeV, the Galactic
Center is within the mean decay length, and above 2~EeV most of the
galactic disk is within range for neutron astronomy.  

\noindent In a previous paper \citep{Blind}, the Pierre Auger
Collaboration published a blind search for a neutron flux from any
point of the sky with declination less than $+15^\circ$, and celestial
maps of flux upper limits were presented.  No point stood out as
statistically significant among the large number of trial source
locations.  In this paper, the search is limited to a small number of
trials, each being a kind of ``stacked analysis'' for a set of
candidate sources from an astrophysical catalog.  The hypothesis is
that many (or all) of the candidate sources of a given class are
indeed emitting neutrons, so the combined signal should be more
significant than that of a single target by itself.  Performing the
analysis on only a small number of {\it target sets} avoids a large
statistical penalty.

\noindent The ``ankle'' of the energy spectrum (at about 5~EeV,~\citep{Spectrum, ICRC2013_S}) may
represent a transition from galactic to extragalactic sources of
cosmic rays \citep{Linsley, Hillas}, although an alternative scenario holds that
the ``dip'' of the ankle is formed by $e^{\pm}$ pair production in a
spectrum of extragalactic protons \citep{dip-model}.  At energies in
the range 1-5~EeV, Auger, HiRes, and Telescope Array have all found 
their data to be consistent with a significant component of protons among 
the cosmic rays \citep {xsection, Xmax, HiRes, TA}. It is of particular interest 
to determine whether or not EeV protons are being produced at discrete
sources in the Galaxy.  A sensitive search for large scale anisotropy
\citep{LS} has found that the dipole anisotropy near 1~EeV is lower
than what would be expected for protons emitted from the disk of the
Galaxy.  However, those expectations necessarily rely on general
properties of the magnetic field in the Galaxy and its halo that have
some uncertainties.  The search for sources of EeV protons in the
Galaxy is still highly motivated.
 
\noindent Any proton source is expected to produce some neutrons due to
pion photo-production and nuclear interactions of the protons near the
source.  The detection of hadronic production of TeV gamma rays in
some galactic sources~\citep{Hadrons} provides direct evidence that the conditions
for neutron production are favorable at least in those sources.  TeV
gamma-ray sources in the Galaxy are among the candidate sources
targeted in this study.  A $1/E^2$ differential energy spectrum of
protons from TeV to EeV in some of the H.E.S.S. sources\footnote{\url{http://www.mpi-hd.mpg.de/hfm/HESS/pages/home/sources}}
 would produce a neutron flux that is readily detectable by Auger.  The
Galactic Center is a special candidate source that is well exposed to
the Auger Observatory.  The results here also update an earlier search for
neutral particles from the Galactic Center~\citep{GC}.

\noindent The production of neutrons with the hadronic production of $\pi^{+}$ mesons 
is necessarily accompanied by photons from decay of similarly-produced 
neutral pions. The Auger Observatory can search for the existence of EeV 
photon fluxes using the hybrid data set that includes air fluorescence measurements. 
The search for point sources of photons near 1 EeV will be reported separately.
%~\citep{Kuempel-paper}.

\section{The data set}
\label{data_set}

The Pierre Auger Observatory~\citep{NIM_EA} is centered at latitude 
$35.3^\circ$~S, longitude $69.3^\circ$~W, near Malarg\"{u}e, Argentina, 
with mean altitude 1400~meters above sea level (870 g cm$^{-2}$ atmospheric 
depth). The completed surface detector array consists of 1660
water-Cherenkov stations covering an area of about 3000 km$^2$ on a
triangular grid with 1.5 km spacing, allowing secondary muons,
electrons, and photons to be sampled at ground level with a duty cycle
of nearly 100\%.

\noindent 
The data set analyzed here consists of events recorded by the surface
detector (SD) from 2004 January 1 to 2013 October 31. Events used in
this analysis have zenith angles less than $60^\circ$.  Moreover, an
event is accepted only if all six nearest neighbors of the station
with the highest signal were operational at the time of the event.
This is the standard geometrical aperture cut that ensures good event
reconstruction \citep{Acceptance}.  Periods of array instability have
been omitted from the data set.  The total exposure of the array with
these cuts is 35,967~km$^2$ sr yr for the period of time analyzed
here, yielding 854,270 events with $E\geq 1$~EeV.

\section{Target sets and weighted target sets}
\label{Targets}

The search is performed on nine target sets of astrophysically
interesting directions. These classes of candidate sources are the
millisecond pulsars~\citep{ATNF}, $\gamma$-ray pulsars~\citep{Fermi},
low-mass X-ray binaries~\citep{LXMB} (LMXB), high-mass X-ray
binaries~\citep{HXMB} (HMXB), H.E.S.S. Pulsar Wind Nebulae, the other H.E.S.S.
identified sources, the H.E.S.S. unidentified sources,
microquasars\footnote{\url{http://www.aim.univ-paris7.fr/CHATY}}, and
magnetars\footnote{McGill Pulsar Group,
\url{http://www.physics.mcgill.ca/~pulsar/magnetar/main.html}}.  In
addition to these target sets, the Galactic Plane and the Galactic
Center are considered as two additional single-element sets for a
total of eleven target sets. In order to have independent target sets,
a source that appears in two or more sets is retained only in the most
exclusive set (smallest set) while removed from the others.  (The
large Galactic Plane target is allowed to contain targets from other
sets.)  Target sets are tested with and without statistical weights.
A weight is assigned to each candidate source in proportion to its
electromagnetic flux (recorded in the catalog), in proportion to its
exposure to the Auger Observatory, and in proportion to the expected
flux attenuation factor due to neutron decay. The flux attenuation
factor is the fraction of emitted neutrons expected to survive decay
from the distance of the candidate source assuming a $1/E^2$ energy
spectrum for emitted neutrons~\footnote{Since the
H.E.S.S. unidentified sources do not have a precise distance measurement,
the flux attenuation factor is not considered for this target set.}.
The target weights are normalized so that their sum is 1 in each target set.

The results without the weights are also reported since they are
independent of the assumption that neutron emissions are proportional
to the low-energy photon emissions and independent of the choice to
make the weights proportional to each of the three factors.

\section{Method}
\label{Method}

Four energy ranges are used: 1~EeV $\leq E <$ 2~EeV (621,375 events),
2~EeV $\leq E <$ 3~EeV (135,444 events), $E \geq$ 3~EeV (97,451
events), as well as $E \geq$ 1 EeV.  The first three are independent
data sets, while the final cumulative data set should give maximum
sensitivity to a flux that extends over the entire energy range. The 
Auger energy scale has a systematic uncertainty of 14\%~\citep{ICRC2013_E}.

The solid angle size for each target is optimized based on the average
angular resolution for its declination and the energy range as
explained in~\citep{Blind}.  The solid angle is a target circle of
radius 1.05 times the angle within which 68\% of neutron arrival
directions from the candidate source should be included after the
event reconstruction.  For the special case of the Galactic Plane,
the signal-to-noise ratio is optimized by a strip centered on galactic
latitude b=0$^{\circ}$ with half thickness $|\rm{b}|<0.93 \psi$, where $\psi$ is
the mean angular resolution along the Galactic Plane (and within the
exposure of the Observatory) for a given energy range.  For the four
energy ranges above, the half thicknesses are, respectively,
$1.28^\circ$, $1.02^\circ$, $0.72^\circ$, and $1.17^\circ$.

To recognize the existence of an excess of events in any target circle 
and any of the energy ranges, it is necessary to know the number that 
is expected in that circle without the neutron flux. The expected number 
of events in a given target circle is taken to be the average number found
 in 10,000 simulated data sets, each having the same number of events as 
in the actual data set. As in~\citep{Blind}, the arrival directions in 
a simulation event set are produced by sampling independently from the 
measured distributions for zenith angle, azimuth angle, and sidereal time. 
 The average of many simulation data sets has no structure on small angular 
scales, providing a robust measure of the expected cosmic-ray background 
in each target.

The result for any target $i$ is summarized by a p-value $p_i$.  This
p-value is here defined by $p_i\equiv \frac{1}{2}
[Poisson(n,b)+Poisson(n+1,b)]$, where $Poisson(n,b)$ is the
probability of getting $n\ or\ more$ arrival directions in the target
when the observed value is $n$, and the expected number from the background 
is $b$, as determined using simulated data sets. Averaging the values 
for $n$ and $n+1$ avoids the bias toward high p-values that occurs 
with $Poisson(n,b)$ and the bias toward low p-values that occurs with 
$Poisson(n+1,b)$ for pure background fluctuations. 
When combining probabilities from a
large target set with the Fisher formula~\citep{Fisher}, it is important that
the individual p-values $p_i$ have uniform expected distributions in
the absence of any signal.  Then for $N$ targets with probabilities
$p_i$, $i=1,2,...,N$, the chance probability for their product ($\Pi$) not to be 
greater than their actual product ($\Pi_{0}$) is:   

\[
\mathbb{P} (\Pi \leq \Pi_{0}) = \Pi_{0} \sum\limits_{j=0}^{N-1} \frac{(-\log\Pi_{0})^{j}}{j!} = 1-Poisson(N,-\log \Pi_{0}).
\]

For a weighted set of $N$ targets with weights $w_i$, the combined
p-value $P_w$ is given by Good's formula~\citep{Good}.  It is the
chance probability for the weighted product of p-values (but sampled
from uniform distributions) not to be greater than the actual weighted
product $\Pi_w$. The weighted product is the product of factors
$p_i^{w_i}$.  Each p-value $p_i$ is raised to the power $w_i$ in the
product of p-values, so the weight $w_i$ can be regarded as the
``number of times'' the result for target $i$ is counted relative to
other targets of the set.  In practice, the Good combined $P_w$ is
evaluated numerically using an ensemble of sets {$p_i$}
($i=1,2,...,N$) with every $p_i$ sampled randomly between 0 and 1.

\section{Results for the target sets}
\label{blind_search}

Results are shown in Tables~\ref{tab:stackAnalysis} and~\ref{tab:topTargets}.  
The first table gives the unweighted combined p-value $P$ for each of the 
11 target sets and for each of the four energy ranges.  The weighted combined 
p-value $P_w$ is also given for each of the 9 target sets that have multiple
targets.  

The second table presents specific information for each target set
about the candidate source that had the smallest individual p-value
$p_i$ for the full energy range E $\geq$ 1 EeV. The direction of the
source is given together with the observed number of events in the
target, the expected number, the neutron flux upper limit, energy flux
upper limit (assuming a $1/E^2$ spectrum), and the p-value. The final
column gives the penalized p-value $p^*=1-(1-p)^N$.  This is the
chance probability that one or more of the $N$ candidate sources in
the target set would have a p-value less than $p$ if each p-value were
randomly sampled from the uniform probability distribution.

The method to compute the limits is the same as the one explained in~\citep{Blind}, 
where the definition of the upper limit in the number of neutrons is that of 
Zech~\citep{Zech}, using a 95\% confidence level.

\begin{table}[ph]
  \footnotesize
  \caption{Results of the combined analysis for each target set and each energy range.}
  \begin{center}
  \begin{tabular}{llllll|llll}
    \hline \hline \\[1ex]
    & & \multicolumn{4}{c}{Weighted P-value $P_w$} & \multicolumn{4}{c}{Unweighted P-value $P$} \\[1ex]
    \cline{3-6} \cline{7-10} \\[1ex]
    Class    & No. &  $\geq$1 EeV & 1-2 EeV & 2-3 EeV & $\geq$3 EeV & $\geq$1 EeV & 1-2 EeV & 2-3 EeV & $\geq$3 EeV \\[1ex]
    \hline \\[1ex]
    % -- Weighted and Unweighted --
    msec PSRs         & 68  & 0.48  & 0.40   & 0.22  & 0.61  & 0.86  & 0.53  & 0.64 & 0.65 \\[1ex]
    $\gamma$-ray PSRs & 77  & 0.23  & 0.13   & 0.71  & 0.24  & 0.82  & 0.96  & 0.38 & 0.64 \\[1ex]
    LMXB              & 87  & 0.37  & 0.43   & 0.81  & 0.40  & 0.041 & 0.12  & 0.13 & 0.54 \\[1ex]
    HMXB              & 48  & 0.014 & 0.011  & 0.061 & 0.27  & 0.095 & 0.090 & 0.22 & 0.66 \\[1ex]
    H.E.S.S. PWN      & 17  & 0.083 & 0.021  & 0.98  & 0.21  & 0.88  & 0.87  & 0.75 & 0.042 \\[1ex]
    H.E.S.S. other    & 16  & 0.91  & 0.93   & 0.94  & 0.35  & 0.42  & 0.83  & 0.66 & 0.028 \\[1ex]
    H.E.S.S. UNID     & 15  & 0.82  & 0.78   & 0.98  & 0.94  & 0.48  & 0.69  & 0.88 & 0.86 \\[1ex]
    Microquasars      & 13  & 0.28  & 0.16   & 0.85  & 0.96  & 0.031 & 0.26  & 0.23 & 0.56 \\[1ex]
    Magnetars         & 16  & 0.69  & 0.52   & 0.60  & 0.46  & 0.73  & 0.85  & 0.83 & 0.41 \\[1ex]
    Gal. Center       & 1   & -     & -      & -     & -     & 0.24  & 0.48  & 0.22 & 0.17 \\[1ex]
    Gal. Plane        & 1   & -     & -      & -     & -     & 0.96  & 0.91  & 0.70 & 0.25 \\[1ex]
    \hline \hline
   % }% 
  \end{tabular}
  \end{center}
  \label{tab:stackAnalysis}
\end{table}

\begin{table}[ph]
  \footnotesize
  \caption{Results for the most significant target from each target set. The upper limits are computed at 95\% confidence level.}
  \begin{center}
    \begin{tabular}{lllllllllll}
      \hline \hline \\[1ex]
      & & & & & Flux U.L. & E-Flux U.L. &  & p-value \\[0.5ex]
      Class & RA [$^\circ$] & Dec [$^\circ$] & Obs & Exp & [\rm{km$^{-2}$ yr$^{-1}$}] & [\rm{eV cm$^{-2}$ s$^{-1}$)}] & p-value & (penalized)\\[1ex]
      \hline \\[1ex]
      msec PSRs         & 260.27 & -24.95 & 237  & 214  & 0.019 & 0.14 & 0.058   & 0.98 \\[1ex]
      $\gamma$-ray PSRs &   8.59 &  -5.58 & 176  & 149  & 0.024 & 0.18 & 0.016   & 0.70 \\[1ex]
      LMXB              & 264.57 & -26.99 & 265  & 219  & 0.028 & 0.20 & 0.0012  & 0.10 \\[1ex]
      HMXB              & 152.45 & -58.29 & 283  & 248  & 0.019 & 0.14 & 0.014   & 0.49 \\[1ex]
      H.E.S.S. PWN      & 128.75 & -45.60 & 275  & 248  & 0.018 & 0.13 & 0.043   & 0.53 \\[1ex]
      H.E.S.S. other    & 269.72 & -24.05 & 235  & 211  & 0.019 & 0.14 & 0.054   & 0.59 \\[1ex]
      H.E.S.S. UNID     & 266.26 & -30.37 & 251  & 227  & 0.018 & 0.13 & 0.055   & 0.57 \\[1ex]
      Microquasars      & 262.75 & -26.00 & 247  & 216  & 0.022 & 0.16 & 0.020   & 0.23 \\[1ex]
      Magnetars         & 81.50  & -66.08 & 268  & 241  & 0.016 & 0.11 & 0.040   & 0.48 \\[1ex]
      Gal. Center       & 266.42 & -29.01 & 234  & 223  & 0.014 & 0.10 & 0.24    & - \\[1ex]
      Gal. Plane        & \multicolumn{2}{l}{$|Gal.\ lat.|<1.17^\circ$} & 16965 & 17197 & 0.077 & 0.56 & 0.96 & - \\[1ex]
      \hline \hline
    \end{tabular}
  \end{center}
  \label{tab:topTargets}
\end{table}

\section{Summary and discussion}
\label{discussion}

None of the candidate source classes tested in this study reveals
compelling evidence for fluxes of EeV neutrons.  Moreover, the minimum p-value
for each target set is not statistically significant when penalized
for the number of targets in the set.  Neither collectively nor
individually is there evidence for EeV neutron fluxes from the
candidate sources that were examined.  

The upper limits on the energy flux from these candidate sources, as
shown in Table~\ref{tab:topTargets}, are below the energy fluxes
detected from TeV gamma-ray sources in the Galaxy.  If such a source
were accelerating protons in the same environment to EeV energies with
the $1/E^2$ dependence expected for Fermi acceleration, then the
energy flux of neutrons in the EeV energy decade would exceed the
energy flux in TeV gamma rays, since neutrons are produced more
efficiently than gamma rays of equal energy.

The upper limit on the neutron flux from the Galactic Plane provides a
stringent constraint on models for continuous production of EeV
protons in the Galaxy.  The emission rate in the disk of the Galaxy
must be sufficient to replace those protons as fast as they escape,
providing an estimate on the required emission rate.  The concomitant
neutron emission rate is model dependent.  It could exceed the proton
emission rate if protons are magnetically bound to the sources and
only the produced neutrons escape, yielding EeV protons by their later
decays.  More likely, however, the neutron luminosity at any fixed
energy is less than the proton luminosity. \textit{Based on an estimate} 
of the proton emission rate, the limits here on
neutron flux from the Galactic Plane can be used to derive a \textit{qualified}
upper limit on this ratio
$\eta=\frac{neutron\ luminosity}{proton\ luminosity}$.  The ratio
$\eta$ is closely related to the average optical depth of the source
regions to escaping EeV protons.  A very low value for $\eta$ 
would imply that protons escape remarkably freely without interacting
with photons or nuclei. 

The expected neutron flux from the Galactic Plane can be estimated
from the luminosity density of protons $\omega$ in the Galactic Plane
together with the unknown ratio $\eta$. The luminosity density $\omega$ is the
rate of proton production per unit area of the Plane.  It is given in
terms of the proton density $\rho$, the thickness $H$ of the cosmic-ray 
disk, and the proton residence time $\tau$, by $\omega=\rho H/\tau$. The
density of protons $\rho$ can be written as $\frac{4\pi}{c} f_p I$,
where $I$ is the cosmic-ray intensity and $f_p$ denotes the fraction
of cosmic rays that are protons. Substituting this expression for
$\rho$ yields: \[\omega= 4\pi f_p I \frac{H}{c\tau}.\] 
\noindent for any given energy range. From any
infinitesimal area element of the Galactic Plane {\rm dA} at distance
$r$, the expected neutron flux is ${\rm dF}= \eta \frac{\omega{\rm dA}}{4\pi
r^2}\exp(-r/D)$, where $D$ is the mean neutron decay distance.
(Note that the Auger Observatory has little exposure to the
Anti-Center direction but good exposure for the Galactic Plane in
directions looking toward inner parts and the far side of the Galaxy, 
so the plane is treated crudely as emitting uniformly from
the exposed annulus out to 20 kpc.)
Integrating over an annulus of the Galactic Plane centered on Earth
gives an expected flux of neutrons: \[F_{GP}=\eta 4\pi f_p I
\frac{H}{c\tau} \int_{R_{min}}^{R_{max}}\frac{\exp(-r/D)}{4\pi
r^2}2\pi r {\rm dr} = \eta 2\pi f_p I \frac{H}{c\tau}
\int_{R_{min}}^{R_{max}}\frac{\exp(-r/D)}{r}{\rm dr}.\] Here $R_{min}$
is the distance at which most of the source disk fits within the angle
(based on the angular resolution) used in collecting the signal from
the Galactic Plane, and $R_{max}$ is the distance to the edge of the
Galaxy.  Roughly, $R_{min} \sim 1 $ kpc and $R_{max} \sim 20 $ kpc. Using 
$D=9.2 E$ kpc (with $E$ measured in EeV units), the final integral on the right has the
following values: 1.97, 2.32, 2.53, and 2.27 respectively for the four
energy ranges 1-2, 2-3, $\geq$3, and $\geq$1~EeV.  Each is somewhat
less than $\ln(20/1)=2.996$, which would be obtained in the limit $D \gg
20$ kpc, and it should be noted that the expected flux has only an
(approximately) logarithmic dependence on $R_{max}/R_{min}$, so the
flux estimates are not terribly sensitive to the particular values
adopted here.  The proton fraction $f_p$ just below the ankle of the
cosmic-ray energy spectrum is believed to be at least $30\%$
\citep{xsection,Xmax}.  A rough estimate for $H/c\tau$ is 0.1, based
on proton escape times and density scale heights using model magnetic
fields in the Galaxy.  Using these estimates, together with
$I~\simeq~29$ (km$^2$ sr yr)$^{-1}$ for $E\geq 1$ EeV, the upper limit
$F_{UL}$ on a neutron flux from the Galactic Plane yields a qualified
upper limit on $\eta$ given by $\eta_{UL}=0.08\ F_{UL}$.  The upper
limit $F_{UL}~\simeq~0.077$ (km$^2$yr)$^{-1}$ in
Table~\ref{tab:topTargets} then implies $\eta_{UL}~\simeq~0.006$, which is 
a significant constraint on models for continuous production of EeV 
protons in the Galaxy.

While the scope of this paper is a search for galactic point sources, it 
should be noted that a flux of neutrons may be produced diffusely at 
a small level by interactions of EeV cosmic-ray protons in the galactic 
disk. Our bound on $\eta$ might therefore be slightly conservative.

A positive detection of neutron sources would have identified sources
of EeV protons in the Galaxy.  The null results here leave open the
question of whether the observed EeV protons are produced in the
Galaxy or whether they fill the space also between galaxies.  They
might be produced in transient events within the Galaxy, like
supernova explosions or rare gamma ray bursts.  The Auger Observatory
could detect a strong flux of neutrons only at times of the bursts.
The non-detection of any neutron flux averaged over the Auger exposure
time does not constrain models for infrequent transient sources of EeV
protons in the Galaxy.  Also, protons emitted in jets would produce
neutron jets with possibly none of them pointing toward Earth.  Models
with transient sources or such jet sources are better constrained by
the absence of any strong anisotropy of the protons themselves
\citep{RA,Dipole}.

\section*{Acknowledgements}
The successful installation, commissioning, and operation of the Pierre Auger Observatory would not have been possible without the strong commitment and effort from the technical and administrative staff in Malarg\"{u}e. 

We are very grateful to the following agencies and organizations for financial support: 
Comisi\'{o}n Nacional de Energ\'{\i}a At\'{o}mica, Fundaci\'{o}n Antorchas, Gobierno De La Provincia de Mendoza, Municipalidad de Malarg\"{u}e, NDM Holdings and Valle Las Le\~{n}as, in gratitude for their continuing cooperation over land access, Argentina; the Australian Research Council; Conselho Nacional de Desenvolvimento Cient\'{\i}fico e Tecnol\'{o}gico (CNPq), Financiadora de Estudos e Projetos (FINEP), Funda\c{c}\~{a}o de Amparo \`{a} Pesquisa do Estado de Rio de Janeiro (FAPERJ), S\~{a}o Paulo Research Foundation (FAPESP) Grants \# 2010/07359-6, \# 1999/05404-3, Minist\'{e}rio de Ci\^{e}ncia e Tecnologia (MCT), Brazil; MSMT-CR LG13007, 7AMB14AR005, CZ.1.05/2.1.00/03.0058 and the Czech Science Foundation grant 14-17501S, Czech Republic;  Centre de Calcul IN2P3/CNRS, Centre National de la Recherche Scientifique (CNRS), Conseil R\'{e}gional Ile-de-France, D\'{e}partement Physique Nucl\'{e}aire et Corpusculaire (PNC-IN2P3/CNRS), D\'{e}partement Sciences de l'Univers (SDU-INSU/CNRS), France; Bundesministerium f\"{u}r Bildung und Forschung (BMBF), Deutsche Forschungsgemeinschaft (DFG), Finanzministerium Baden-W\"{u}rttemberg, Helmholtz-Gemeinschaft Deutscher Forschungszentren (HGF), Ministerium f\"{u}r Wissenschaft und Forschung, Nordrhein Westfalen, Ministerium f\"{u}r Wissenschaft, Forschung und Kunst, Baden-W\"{u}rttemberg, Germany; Istituto Nazionale di Fisica Nucleare (INFN), Ministero dell'Istruzione, dell'Universit\`{a} e della Ricerca (MIUR), Gran Sasso Center for Astroparticle Physics (CFA), CETEMPS Center of Excellence, Italy; Consejo Nacional de Ciencia y Tecnolog\'{\i}a (CONACYT), Mexico; Ministerie van Onderwijs, Cultuur en Wetenschap, Nederlandse Organisatie voor Wetenschappelijk Onderzoek (NWO), Stichting voor Fundamenteel Onderzoek der Materie (FOM), Netherlands; National Centre for Research and Development, Grant Nos.ERA-NET-ASPERA/01/11 and ERA-NET-ASPERA/02/11, National Science Centre, Grant Nos. 2013/08/M/ST9/00322 and 2013/08/M/ST9/00728, Poland; Portuguese national funds and FEDER funds within COMPETE - Programa Operacional Factores de Competitividade through Funda\c{c}\~{a}o para a Ci\^{e}ncia e a Tecnologia, Portugal; Romanian Authority for Scientific Research ANCS, CNDI-UEFISCDI partnership projects nr.20/2012 and nr.194/2012, project nr.1/ASPERA2/2012 ERA-NET, PN-II-RU-PD-2011-3-0145-17, and PN-II-RU-PD-2011-3-0062, the Minister of National  Education, Programme for research - Space Technology and Advanced Research - STAR, project number 83/2013, Romania; Slovenian Research Agency, Slovenia; Comunidad de Madrid, FEDER funds, Ministerio de Educaci\'{o}n y Ciencia, Xunta de Galicia, Spain; The Leverhulme Foundation, Science and Technology Facilities Council, United Kingdom; Department of Energy, Contract No. DE-AC02-07CH11359, DE-FR02-04ER41300, and DE-FG02-99ER41107, National Science Foundation, Grant No. 0450696, The Grainger Foundation, USA; NAFOSTED, Vietnam; Marie Curie-IRSES/EPLANET, European Particle Physics Latin American Network, European Union 7th Framework Program, Grant No. PIRSES-2009-GA-246806; and UNESCO.

\end{document}